# X-ray studies of radio-silent neutron stars in the remnants of supernova explosions.

Collazos J. Alfredo[1]

[1]*St. Petersburg State Polytechnical University, Politechnicheskaya 29, 195251, St. Petersburg, Russia*

**ABSTRACT**

The present work is devoted to the study of the supernova remnant and its neutron star XMMU J172054. 5-372652. The paper analyzes observations in 2009 and 2018, data for this system obtained by the Chandra x ray observatory. The analysis showed that the neutron star G350. 1-0. 3 with M = 1, 4 Msun, R = 13 MS, dist = 4, 5 KPC, we also measured its temperature in two groups: the first group, including observations made in 2009, and the second group, including 5 observations made at different times of the year 2018.

In addition, in this work, we study all important aspects of the supernova remnants of G350. 1-0. 3, and obtain their spectra using the collided plasma, nonequilibrium, constant temperature (VNEI) model, which assumes a constant temperature and a single ionization parameter. It gives a characteristic of the spectrum, but is not a physical model, we will precede the spectra obtained using this model.

## 1 INTRODUCTION

Supernovae play an important role in modern astrophysics. They are vital for the chemical evolution of the universe, and are also one of the most important sources of energy for the interstellar medium. Part of this energy is in the form of cosmic rays, which have an energy density of 1-2 eV $cm^{-3}$ in the Galaxy, which is thus about a third of the total energy density of the interstellar medium. Finally, supernovae, especially Type Ia supernovae, play a central role in modern cosmology, since their brightness allows them to be detected at high redshifts (Paolo & Heavens 2007). their use led to the recognition that the expansion of the universe is accelerating, not slowing down (Riess & Filippenko 2007).

G350. 1-0. 3 is a bright radio source in our Galaxy. Its linear polarization and non-thermal spectrum initially led to its classification as SNR (Canizares & Neighbours 1975), but the high-resolution image revealed a distorted and elongated morphology, very different from the shell structure usually observed in such sources.

It was later shown that G350. 1-0. 3 is a very young and bright SNR expanding into a dense surrounding gas. An associated neutron star has been discovered in it, which is probably an addition to the growing class of "central compact objects" (CCOs) in the remnants of supernovae.

Many supernova remnants (SNRs) expand into complex environments formed by molecular clouds and strong winds from OB associations. If our understanding of SNR is to progress, we must strive to identify and understand these complex systems.

The subject of the study is to obtain the spectra of observations of Chandra G350. 1-0.3, to analyze the models most suitable for their study, using the tools xspec, sherpa.

The aim of the work is to analyze the X-ray observations of the Chandra Observatory conducted in 2009-2018 and to choose a spectral model that best describes the spectrum of the CCO.

[1] E-mail: jacollazos@utp.edu.co

## 2 CHANDRA OBSERVATIONS

G350. 1-0. 3 is a supernova remnant located in the constellation Scorpio. It is located in the Milky Way and is possibly associated with a neutron star (XMMU J172054.5-372652) formed as a result of the same supernova explosion. Previously, this object was mistakenly classified as a distant galaxy (Witt A. 2002).

G350. 1-0. 3, a bright radio source inside the Milky Way, was initially identified by comparing observations from the Molonglo Observatory Synthesis Telescope and the Parks Radio Telescope and classified as a supernova remnant in publications in 1973 and 1975.

However, later higher-resolution images in the mid-80s revealed an unexpected irregular morphology that was significantly different from other sources of supernova remnants. Then it was claimed that G350. 1-0. 3 was a radio galaxy or galaxy cluster, which led to a reclassification, as a result of which the catalogs of supernova remnants downgraded the object to a "candidate for a supernova remnant" or completely excluded it; G350.1-0.3 was subsequently "forgotten"(Borkowski & Miltich 2020).

A study published in 2008 combined archived data and new images from the European Space Agency's XMM-Newton orbiting X-ray telescope to demonstrate that G350. 1-0.3 is a supernova remnant. The researchers determined that the strange shape of the object arose as a result of an explosion near a dense gas cloud, about 15,000 light-years from Earth, which prevented uniform expansion and led to its elongated shape.

The researchers also determined that the nearby thermal X-ray source XMMU J172054. 5-372652 is a central compact supernova object (Borkowski & Miltich 2020).

G350. 1-0. 3 is eight light-years across and about 900-1000 years old, making it one of the youngest and brightest supernova remnants in the Milky Way. It is unlikely that people would have seen a supernova explosion, because the intermediate interstellar dust would probably have prevented its observation from Earth.

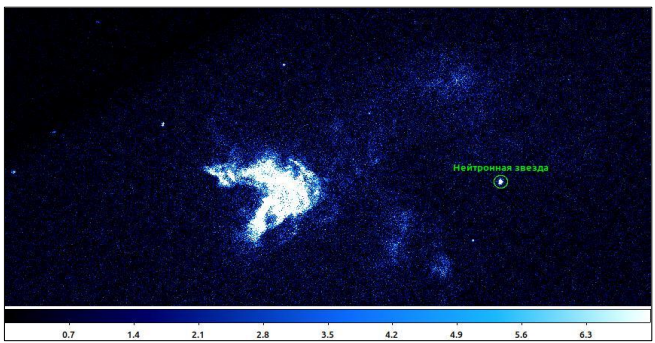

**Fig 1.** Supernova remnant G350. 1-0. 3 and neutron star XMMU J172054. 5-372652

The Chandra X-Ray Observatory observed G350. 1-0.3 on May 21, 2009 and July 5, 2018. The total time of the two observations is 82.97 kps and 189.24 kps, respectively. More detailed information about individual observations can be seen in Table 1. The data is processed using chandra repro in CIAO 4.13, and the images are obtained using Ds9 provided by the Chandra X-ray center.

| Name | Observatory | observations | Year |
|---|---|---|---|
| G350.1-0.3 | Chandra | 10102 | 2009 |
| | | 20312 | 2018 |
| | | 20313 | |
| | | 21118 | |
| | | 21119 | |
| | | 21120 | |

**Table 1.** Chandra Observation for G350. 1-0. 3

The ACIS-S tool is designed for detecting spectra and worked in the standard mode and continuous data accumulation.

G350. 1-0. 3 was observed on May 21, 2009 at 05: 11: 23, and five segments in July 2018 for a total of 189. The average distance between the two epochs is 9.126 years. In all observations, $G$350.1-0.3 was placed on an advanced ACIS S3 CCD spectrometer that operated in the weak source mode.

Looking at the X-ray image of G350. 1-0. 3 taken in 2009, we see in Figure 2. there is practically no X-ray radiation to the West, and XMMU J172054.5-372652 is observed, which is our neutron star, our CCO. The intensity of X-rays is higher in the eastern region than in the Northern and Southern regions. XMMU J172054. 5-372652 is located in the western region, which is about 20 times weaker than the eastern one.

According to the results of the work (Borkowski & Miltich 2020), it can be said that some weaker areas have more or less sunny abundance. The proper motions of the expansion show that G350. 1-0.3 is eight light-years wide and about 600 years old, regardless of the distance. This is the third youngest known supernova remnant in the Galaxy and one of the most asymmetric.

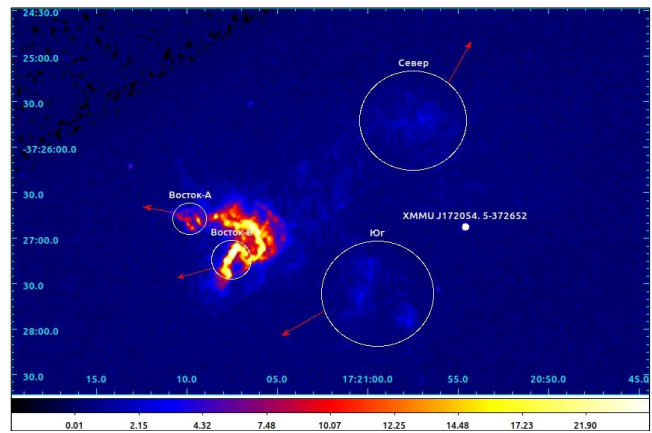

**Fig 2.** Chandra ACIS-S image taken in 2009. The energy range is 0.5-7.0 keV.

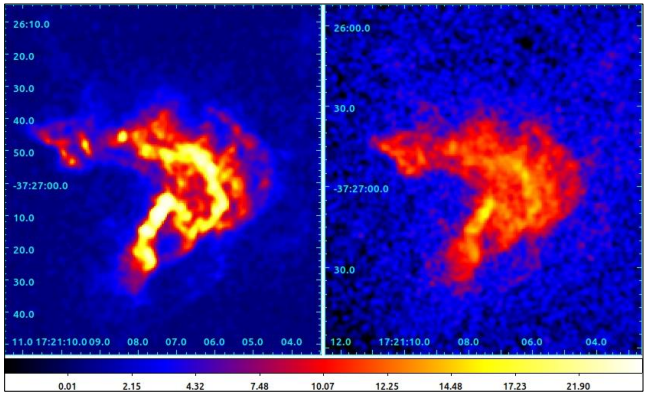

**Fig 3.** Images of Chandra G350. 1-0. 3, on the left we see a supernova remnant in 2009, on the right we see a supernova remnant in 2018. The figures are not shown to the same scales along the flow.

Corresponding areas of motion analysis, using the maximum likelihood method. The arrows indicate the relative value of the 2D velocities and the direction of movement of the emissions.

Next, we will analyze the spectra of the selected areas, we can do this using specextract. To select the spectrum of the southern region, we use the manual and perform the processing as shown below.

```
(ciao-4.13) G350.1-0.3/10102/repro/Region1$ pset specextract
infile="acisf10102_repro_evt2.fits[sky=region(src.reg)]"
(ciao-4.13) G350.1-0.3/10102/repro/Region1$ pset specextract
bkgfile="acisf10102_repro_evt2.fits[sky=region(bkg.reg)]"
(ciao-4.13) G350.1-0.3/10102/repro/Region1$ pset specextract outroot=10102
(ciao-4.13) Desktop/G350.1-0.3/10102/repro/Region1$ pset specextract correctpsf=yes
(ciao-4.13) Desktop/G350.1-0.3/10102/repro/Region1$ pset specextract weight=no
(ciao-4.13) Desktop/G350.1-0.3/10102/repro/Region1$ specextract
```

Now we can get the spectrum using Sherpa as follows:

```
Sherpa 4.13.0
Python 3.8.2 (default, Mar 25 2020, 17:03:02)
IPython profile: sherpa
Using matplotlib backend: Qt5Agg
sherpa In [1]: load_pha("10102.pi")
sherpa In [2]: plot_data()
```

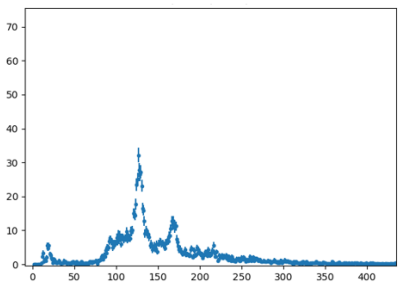

**Fig 4.** The spectrum of the northern region, we used the data process in specextract and Sherpa to obtain this spectrum.

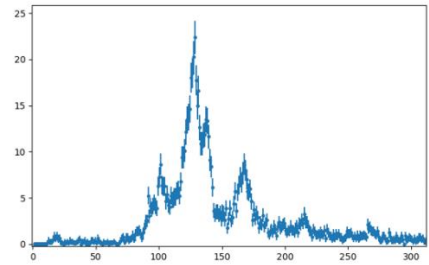

**Fig 5.** The spectrum of the Eastern A region.

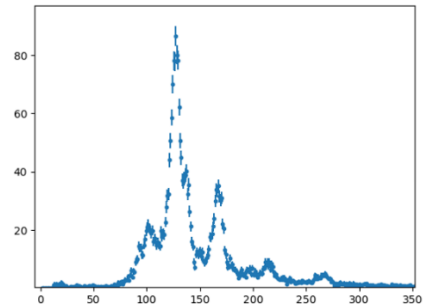

**Figure 6.** The spectrum of the Eastern B region.

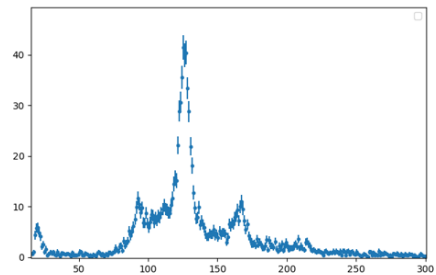

**Figure 7.** Spectrum of the Southern region.

After obtaining the spectra of the selected regions, we can conclude that the most intense spectra belong to the eastern regions, in addition, analyzing the spectra selected in 2009 and 2018.

## 3 DATA PROCESSINDG AND MODELS

In this chapter, we present an original result of analyzing the spectra of a Central Compact Object in the supernova remnant G350.1-0.3. Due to the fact that the supernova remnant is a young object, we can assume that the SSO may be cooling. To verify this, we analyzed Chandra's data for 2009 and 2018.

For the object of research G350. 1-0. 3, it was decided to use the data provided by the Chandra x-ray observatory, below we can see the steps necessary for selecting data, obtaining images of 2009 and 2018. To get the observations of Chandra, we must first have downloaded the files from the Chandra Data Archive. For our particular case G350. 1-0. 3, we find 6 observations (Table 1).

Let's continue downloading all 6 files to study, from the Ubuntu terminal, we need to open the files individually.

Conda is an open source package management system and environment management system that runs on Windows, macOS and Linux. Conda quickly installs, launches, and updates packages and their dependencies. Conda easily creates, saves, loads, and switches between environments on your local computer. It was created for Python programs, but it can package and distribute software for any language.

The Chandra_repro processing command automates the recommended data processing steps presented in the CIAO analysis streams. The command reads the primary data and creates a new bad pixel file, a new level=2 event file, and a new fits file. Chandra_repro can be used to re-process any ACIS and HRC images or grid data. For the 2009 and 2018 observations, we receive files after processing using chandra_repro, which will serve us to select the desired spectra.

After obtaining the spectra and analyzing the data process, now let's study 3 different spectral models for our neutron star, we will describe each model, and also look at the process in xspec, data input and the final result. The selected models are as follows:

- The XSPEC bbodyrad model: a blackbody spectrum, a normalized region.
- The XSPEC carbatm model: The non-magnetic carbon atmosphere of a neutron star.
- The XSPEC nsx model: a neutron star with a non-magnetic atmosphere.

**The carbatm model**

Carbatm: The non-magnetic carbon atmosphere of a neutron star. The model provides the spectra emitted by the non-magnetic carbon atmosphere of a neutron star. The spectra of the model in the range from 0 to 20 keV of photon energy (displacement) are calculated in the grid of accelerations of the gravity of the surface log (g) = 13.7 - 14.9 (in CGS units) and effective temperatures T = 1 - 4 MK. For a given set of parameters, the gravity of the surface gy with a volume up to the Red gravitational z results in the mass M and radius R of the star. The test spectra are calculated by linear interpolation between the spectra of the nearest model in the T-log (g) grid and the boundaries of energy deposits. The number of photons in each container is divided by (1 + z).

Parameters:

- Effective temperature T (MK)
- Neutron star mass M
- Neutron star radius R (km)
- Normalization
- Of the source in units of 10 kpc, and A characterizes the fraction of the surface emitting radiation (A = 1 corresponds to the case when radiation comes from the entire surface).

XSPEC is a command-driven interactive program for approximating the X-ray spectrum, designed to be completely independent of the detector, so it can be used with any spectrometer. XSPEC was used to analyze data from HEAO-1 A2, the Einstein Observatory, EXOSAT, Ginga, ROSAT, BBXRT, ASCA, CGRO, IUE, RXTE, Chandra, XMM-Newton, Integral / SPI, Fermi, Swift, Suzaku, NuSTAR, and Hitomi. Now there are more than 9300 articles in ADS that mention XSPEC.

Now, once the model is defined, and having hidden parameters, we can use the sxpec tool to get the model applied to our case, we choose the following:

```
                    XSPEC version: 12.11.1
          Build Date/Time: Tue Jun  1 20:10:13 2021

========================================================

Model TBabs<1>*carbatm<2> Source No.: 1  Active/On
Model Model Component Parameter Unit   Value
 par  comp
          Data group: 1
 1   1   TBabs    nH     10^22  6.34179  +/- 5.67713E-02
 2   2   carbatm  T      MK     2.01690  +/- 6.45997E-03
 3   2   carbatm  NSmass Msun   1.40000   frozen
 4   2   carbatm  NSrad  km     13.0000   frozen
 5   2   carbatm  norm          4.93800   frozen
          Data group: 2
 6   1   TBabs    nH     10^22  6.34179   = p1
 7   2   carbatm  T      MK     1.98809  +/- 5.24677E-03
 8   2   carbatm  NSmass Msun   1.40000   = p3
 9   2   carbatm  NSrad  km     13.0000   = p4
10   2   carbatm  norm          4.93800   = p5
________________________________________________________

Using energies from responses.
Fit statistic : Chi-Squared          120.36    using 111 bins.
                Chi-Squared           75.97    using  54 bins.
                Chi-Squared           28.04    using  29 bins.
                Chi-Squared           57.53    using  61 bins.
                Chi-Squared           65.39    using  65 bins.
                Chi-Squared           50.39    using  52 bins.
Total fit statistic                  397.68    with 369 d.o.f.

Test statistic : Chi-Squared         397.68    using 372 bins.

Null hypothesis probability of 1.46e-01 with 369 degrees of freedom
XSPEC12>error 2
 Parameter  Confidence Range (2.706)
    2   2.00642   2.02742  (-0.0104833,0.0105189)
XSPEC12>error 7
 Parameter  Confidence Range (2.706)
    7   1.97899   1.99711  (-0.00909123,0.0090261)
========================================================
```

When setting up the carbatm model, it is necessary to fix the mass and radius of the neutron star, as well as the distance to it. In our case, this is M = 1.4 Msun, R = 13 km, dist = 4.5 KPC.

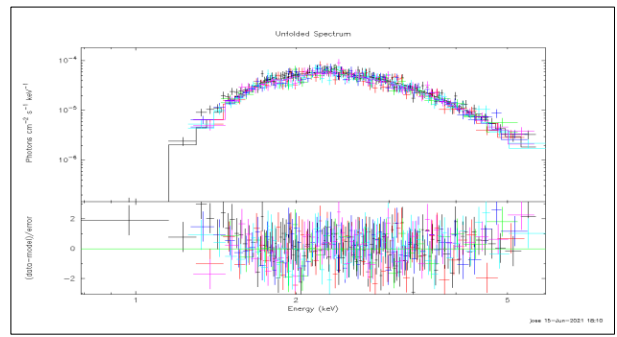

Fig 8. Carbatm graph for a neutron star

In Figure 8. we see the spectrum of XMMU J172054. 5-372652 next to its setting (continuous lines): tbabs*carbatm, green shows the extracted spectrum, red, black, light blue, dark blue, purple, 5 observers in 2018, respectively.

**The nsx model**

The nsx model is interpolated from the grid of atmospheric spectra of neutron stars (NS) to obtain the final spectrum, which depends on the parameters listed below. Atmospheric spectra are obtained using opacity tables calculated by The Opacity Project and are intended for non-magnetic atmospheres (note that nsx is fully compatible with the nsmaxg magnetic atmosphere spectral tables, and the nsx and nsmaxg spectral tables can easily be made compatible with other XSPECS. Suitable NS models). Atmospheric models are constructed by solving the radiation transfer equation, and it is assumed that the atmosphere is in radiation and hydrostatic equilibrium. The atmospheric models depend on the effective surface temperature and surface gravity, where is the gravitational redshift, and is the mass and radius of the NZ, respectively. The parameters are as follows:

Parameters:
- Surface (unbiased) effective temperature
- the gravitational mass of a neutron star (in units of the mass of the Sun).
- R is the radius of the neutron star (in km).
- d the distance to the neutron star (in kpc).
- Switch indicating the model to use
- 1, normalization (although not quite correct, it can be varied to change the size of the radiation área.

Now, once the model is defined, and having hidden parameters, we can use the xspec tool to get the model applied to our case, we choose the following:

```
        XSPEC version: 12.11.1
        Build Date/Time: Tue Jun  1 20:10:13 2021
===================================================
Model TBabs<1>*nsx<2> Source No.: 1  Active/On
Model Model Component  Parameter  Unit     Value
 par  comp
              Data group: 1
  1   1   TBabs    nH      10^22   6.46523   +/- 5.59301E-02
  2   2   nsx      logTeff  K      6.30881   +/- 1.25622E-03
  3   2   nsx      M_ns     Msun   1.40000   frozen
  4   2   nsx      R_ns     km     13.0000   frozen
  5   2   nsx      dist     kpc    4.50000   frozen
  6   2   nsx      specfile        6         frozen
  7   2   nsx      norm            1.00000   frozen
              Data group: 2
  8   1   TBabs    nH      10^22   6.46523   = p1
  9   2   nsx      logTeff  K      6.30254   +/- 1.01166E-03
 10   2   nsx      M_ns     Msun   1.40000   = p3
 11   2   nsx      R_ns     km     13.0000   = p4
 12   2   nsx      dist     kpc    4.50000   = p5
 13   2   nsx      specfile        6         = p6
 14   2   nsx      norm            1.00000   = p7
__________________________________________________

Fit statistic : Chi-Squared        125.14   using 111 bins.
               Chi-Squared          75.53   using  54 bins.
               Chi-Squared          28.85   using  29 bins.
               Chi-Squared          58.60   using  61 bins.
               Chi-Squared          67.68   using  65 bins.
               Chi-Squared          50.83   using  52 bins.
Total fit statistic                406.64   with 369 d.o.f.
Test statistic : Chi-Squared       406.64   using 372 bins.
 Null hypothesis probability of 8.61e-02 with 369 degrees of freedom
XSPEC12>error 9
 Parameter   Confidence Range (2.706)
    9    6.30041    6.30435   (-0.00212662,0.0018169)
XSPEC12>error 2
 Parameter   Confidence Range (2.706)
    2    6.30639    6.31109   (-0.00242179,0.00227566)
XSPEC12>plot ufspec delchi
__________________________________________________
```

When setting up the nsx model, it is necessary to fix the mass and radius of the neutron star, as well as the distance to it. In our case, this is M = 1.4 Msun, R = 13 km, dist = 4.5 KPC, norm= 4.938, A=1. it is necessary to calculate the normalization.

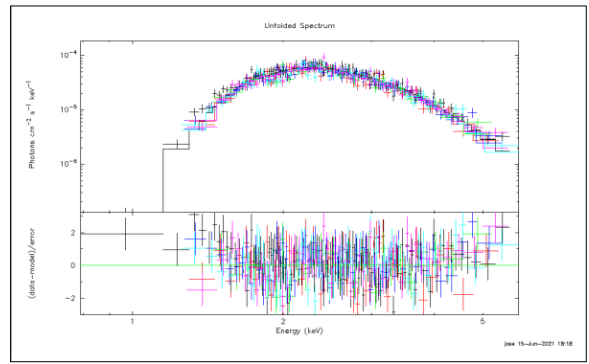

Fig 9. Nsx graph for a neutron star

In Figure 9, we see the spectrum of XMMU J172054. 5-372652 next to its setting (continuous lines): tbabs*nsx, green shows the extracted spectrum, red, black, light blue, dark blue, purple, 5 observers in 2018, respectively.

**The dobyrad model**

The spectrum of the black body. The spectrum of a blackbody with normalization proportional to the surface area.

Now, once the model is defined, and having hidden parameters, we can use the xspec tool to get the model applied to our case, we choose the following:

```
        XSPEC version: 12.11.1
        Build Date/Time: Tue Jun  1 20:10:13 2021

Model TBabs<1>*bbodyrad<2> Source No.: 1  Active/On
Model Model Component  Parameter  Unit     Value
 par  comp
              Data group: 1
  1   1   TBabs     nH     10^22   4.51112   +/- 0.120124
  2   2   bbodyrad  kT     keV     0.518003  +/- 6.36027E-03
  3   2   bbodyrad  norm           2.11915   +/- 0.170306
```

```
                 Data group: 2
  4  1  TBabs      nH      10^22    4.92844   +/- 0.117960
  5  2  bbodyrad   kT      keV      0.517066  +/- 6.02272E-03
  6  2  bbodyrad   norm             2.11915   = p3
________________________________________________________

   Fit statistic : Chi-Squared        104.56   using 111 bins.
             Chi-Squared               83.07   using 54 bins.
             Chi-Squared               27.07   using 29 bins.
             Chi-Squared               58.04   using 61 bins.
             Chi-Squared               61.99   using 65 bins.
             Chi-Squared               52.77   using 52 bins.
   Total fit statistic                387.50   with 367 d.o.f.

   Test statistic : Chi-Squared       387.50   using 372 bins.
   Null hypothesis probability of 2.21e-01 with 367 degrees of freedom

XSPEC12>error 2
 Parameter  Confidence Range (2.706)
    2   0.507246   0.528805  (-0.0107298,0.0108296)
XSPEC12>error5
 Parameter  Confidence Range (2.706)
    5   0.506888   0.527272  (-0.0101471,0.0102363)
```

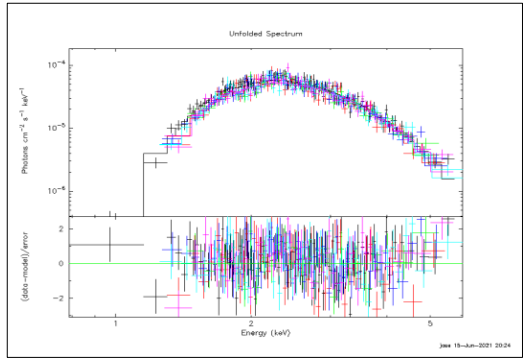

Fig 10. bodyrad graph for a neutron star

In Figure 10. we see the spectrum of XMMU J172054. 5-372652 next to its setting (continuous lines): tbabs*bbodyrad, green shows the extracted spectrum, red, black, light blue, dark blue, purple, 5 observers in 2018, respectively.

| Model   | Chi-Square | dof   |
|---------|------------|-------|
| Nsx     | 406.69     | 1.102 |
| Bodyrad | 387.50     | 1.202 |
| Carbatm | 397.68     | 1.077 |

**Table 2.** The chi-square of each model

| Model   | T(MK)                        | Data |
|---------|------------------------------|------|
| Nsx     | $T_1 = 2.037 \pm 0.536^{-3}$ | 2009 |
|         | $T_2 = 2.004 \pm 0.758^{-3}$ | 2018 |
| Bodyrad | $T_1 = 2.016 \pm 6.45e^{-3}$ | 2009 |
|         | $T_2 = 1.988 \pm 5.24\,e^{-3}$ | 2018 |
| Carbatm | $T_1 = 6.03453 \pm 0.073$    | 2009 |
|         | $T_2 = 5.9993 \pm 0.068$     | 2018 |

**Table 3.** Change in neutron star temperature

we see the results for the temperature of the star obtained by three models. As you can see, the bodyrad temperature is about three times higher than the next atmospheric model. However, this model leads to a radius of the radiating region of about 800 m, which is much smaller than the radius of a neutron star. In this case, it would be possible to observe pulsations of radiation with the period of rotation of the star, which have not yet been detected, so this model can be discarded for the time being.

In the results of atmospheric approximations, which assume radiation from the entire surface of the star, one can see a very small decrease in temperature beyond the error limits, which may indicate that the neutron star may be cooling. At the same time, the carbon atmosphere statistically describes the data a little better than the hydrogen atmosphere. A similar situation is observed for the MTR in the supernova remnant Cas A.

For the case of a carbon atmosphere, the temperature change between two epochs is $0.028 \pm 0.008$ MK.

## 4 RESULTS AND DISCUSSION

From the results of the spectral analysis of the CCO, it can be concluded that, apparently, the neutron star is cooling. However, the temperature change is detected only at the level of 3-sigma. If this is confirmed by further observations, it will be the second young neutron star after Cas A, cooling in real time. Then it can be used to study the processes of neutrino cooling in a young neutron star. The measured surface temperature corresponds to the predictions of neutron star cooling scenarios for an age of about 600 years.

## 5 CONCLUSION

The X-ray emission spectra of a neutron star in the young supernova remnant G350.1-0.3 obtained by the Chandra Observatory in 2009 and 2018 are analyzed.

The radiation is purely thermal and comes from the surface of a neutron star. Its spectrum can be described by three models, including the black body radiation spectrum and the emission spectra of the carbon atmospheres of neutron stars.

Statistically, the most acceptable model is the black body radiation model. It leads to the size of the radiation region of about 800 m, which is much smaller than the radius of the neutron star and should lead to X-ray pulsations with the rotation period of the neutron star, which have not yet been detected.

Atmospheric models assume radiation from the entire surface of the star. Of these, carbatm is statistically the most preferable.

In the latter case, it was found that in the 9 years between observations, the star cooled by $28,000 \pm 8,000$ K. In case of further confirmation of this, it becomes the second, after the star in the Cas A remnant, whose cooling can be studied in real time.